\documentclass[12pt,letterpaper]{article}
\pdfoutput=1

\usepackage{color}
\usepackage{amsfonts,amssymb,amsmath}
\usepackage{MnSymbol,wasysym}
\usepackage{esint}

\usepackage{graphicx}
\usepackage{caption}
\usepackage{subcaption}
\usepackage{booktabs}

\usepackage{listings}

\usepackage{hyperref}

\numberwithin{equation}{section}

\newcommand{\eqn}[1]{(\ref{#1})}
\newcommand{\be}{\begin{equation}}
\newcommand{\ee}{\end{equation}}
\newcommand{\ben}{\begin{displaymath}}
\newcommand{\een}{\end{displaymath}}
\newcommand{\bea}{\begin{eqnarray}}
\newcommand{\eea}{\end{eqnarray}}
\newcommand{\bean}{\begin{eqnarray*}}
\newcommand{\eean}{\end{eqnarray*}}

\newcommand{\bra}[1]{\mbox{$\langle #1 |$}}
\newcommand{\ket}[1]{\mbox{$| #1 \rangle$}}

\newcommand{\eg}{{\it e.g.}}
\newcommand{\ie}{{\it i.e.}}

\newcommand{\commentout}[1]{}

\usepackage{amsmath}
\DeclareMathOperator\atanh{atanh}

\newcommand{\beq}{\begin{equation}}
\newcommand{\eeq}{\end{equation}}
\newcommand{\beqr}{\begin{displaymath}}
\newcommand{\eeqr}{\end{displaymath}}
\newcommand{\beqa}{\begin{eqnarray}}
\newcommand{\eeqa}{\end{eqnarray}}
\newcommand{\beqar}{\begin{eqnarray*}}
\newcommand{\eeqar}{\end{eqnarray*}}

\newcommand{\non}{\nonumber}

\newcommand{\half}{\ensuremath{\frac{1}{2}}}

\newcommand{\cotan}{\ensuremath{\mbox{cotan}}}

\renewcommand{\Re}{\ensuremath{\mathrm{Re}}}
\renewcommand{\Im}{\ensuremath{\mathrm{Im}}}

\begin{document}

\title{\LARGE \bf A note on the S-matrix bootstrap for the 2d O(N) bosonic model}

\author{
	 Yifei He$^1$,
	 Andrew Irrgang$^{1,2}$,
	 Martin Kruczenski$^1$\thanks{E-mail: \texttt{he163@purdue.edu, airrgang1@ivytech.edu, markru@purdue.edu.}} \\
	$^1$ Dep. of Physics and Astronomy, \\ Purdue University, W. Lafayette, IN  \\
	$^2$ Ivy Tech Community College, Lafayette, IN }

\maketitle

\begin{abstract}
 In this work we apply the S-matrix bootstrap maximization program to the 2d bosonic $O(N)$ integrable model which has $N$ species of scalar particles of mass $m$ and no bound states. Since in previous studies theories were defined by maximizing the coupling between particles and their bound states, the main problem appears to be to find what other functional can be used to define this model. Instead, we argue that the defining property of this integrable model is that it resides at a vertex of the convex space determined by the unitarity and crossing constraints. Thus, the integrable model can be found by maximizing any linear functional whose gradient points in the general direction of the vertex, namely within a cone determined by the normals to the faces intersecting at the vertex. This is a standard problem in applied mathematics, related to semi--definite programming and solvable by fast available numerical algorithms. The information provided by the numerical solution is enough to reproduce the known analytical solution without using integrability, namely the Yang-Baxter equation. This situation seems quite generic so we expect that other theories without continuous parameters can also be found by maximizing linear functionals in the convex space of allowed S-matrices. 
\end{abstract}

\clearpage
\newpage



\section{Introduction}
\label{intro}
\subsection{The S-matrix bootstrap maximization program}

  Recently new insights were found in the old idea \cite{Smatrix} of determining the S-matrix directly from its analytic structure, symmetries, crossing and unitarity. This is known to be possible in two dimensional integrable theories but only after using the factorization constraint. More precisely, the Yang-Baxter equation is used to determine an initial, sometimes called minimal, solution. 
  Without using the Yang-Baxter equation, those constraints are not enough to completely determine the S-matrix. However, recently it was found that maximizing the coupling between particles and their bound states led to well-known theories such as a subsector of the sine-Gordon model \cite{Paulos:2016fap,Paulos:2016but}. The physical argument is that, when the spectrum of bound states is fixed, there is a limit on the value of the coupling since stronger couplings will lead to more bound states. This is a very powerful idea, namely that certain theories lay at particular points of the space of allowed theories (or S-matrices) and that those particular points can be found by maximizing certain functionals in that space. Similar arguments lead to bounds on couplings in higher dimensional theories \cite{Paulos:2017fhb} and on couplings to resonances \cite{Miro}.
 
  Motivated by this, we consider the two dimensional bosonic $O(N)$ model as a test of the S-matrix bootstrap maximization program. The main reason is that this model does not have any bound states so the idea of maximizing a coupling does not seem directly applicable and it appears that a new type of functional is required. Also, the S-matrix of this model is exactly known from the work of Zamolodchikov and Zamolodchikov \cite{Zamolodchikov:1978xm} and therefore it is easy to check the results. What we argue in this paper is that the defining property of the integrable model is that it is at a vertex of the convex space of allowed S-matrices and therefore there is an infinite number of linear functionals that are maximized there. Before proceeding to describe the $O(N)$ model, let us mention that in  Appendix B the previous results for the sine-Gordon model are described in relation to the ideas of this paper. 
 
\subsection{The 2d O(N) bosonic model, general properties} 
\begin{figure}
	\centering
	\begin{subfigure}{.75\textwidth}
		\centering
		\includegraphics[trim=0cm 3cm 0cm 3cm, clip=true, width=\linewidth]{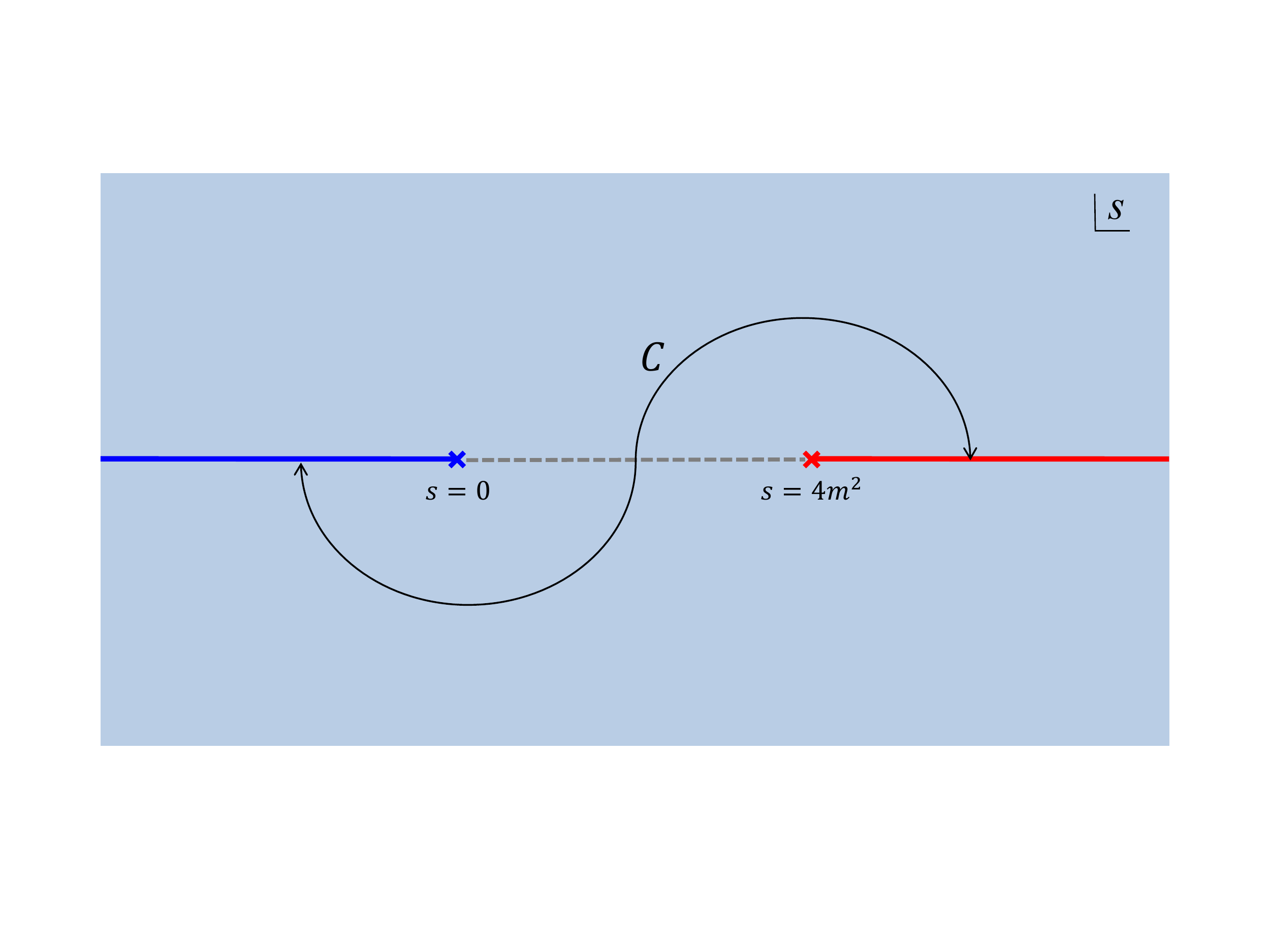}
		\label{splane}
	\end{subfigure} 
	\begin{subfigure}{.75\textwidth}
		\centering
		\includegraphics[trim=0cm 3cm 0cm 3cm, clip=true, width=\linewidth]{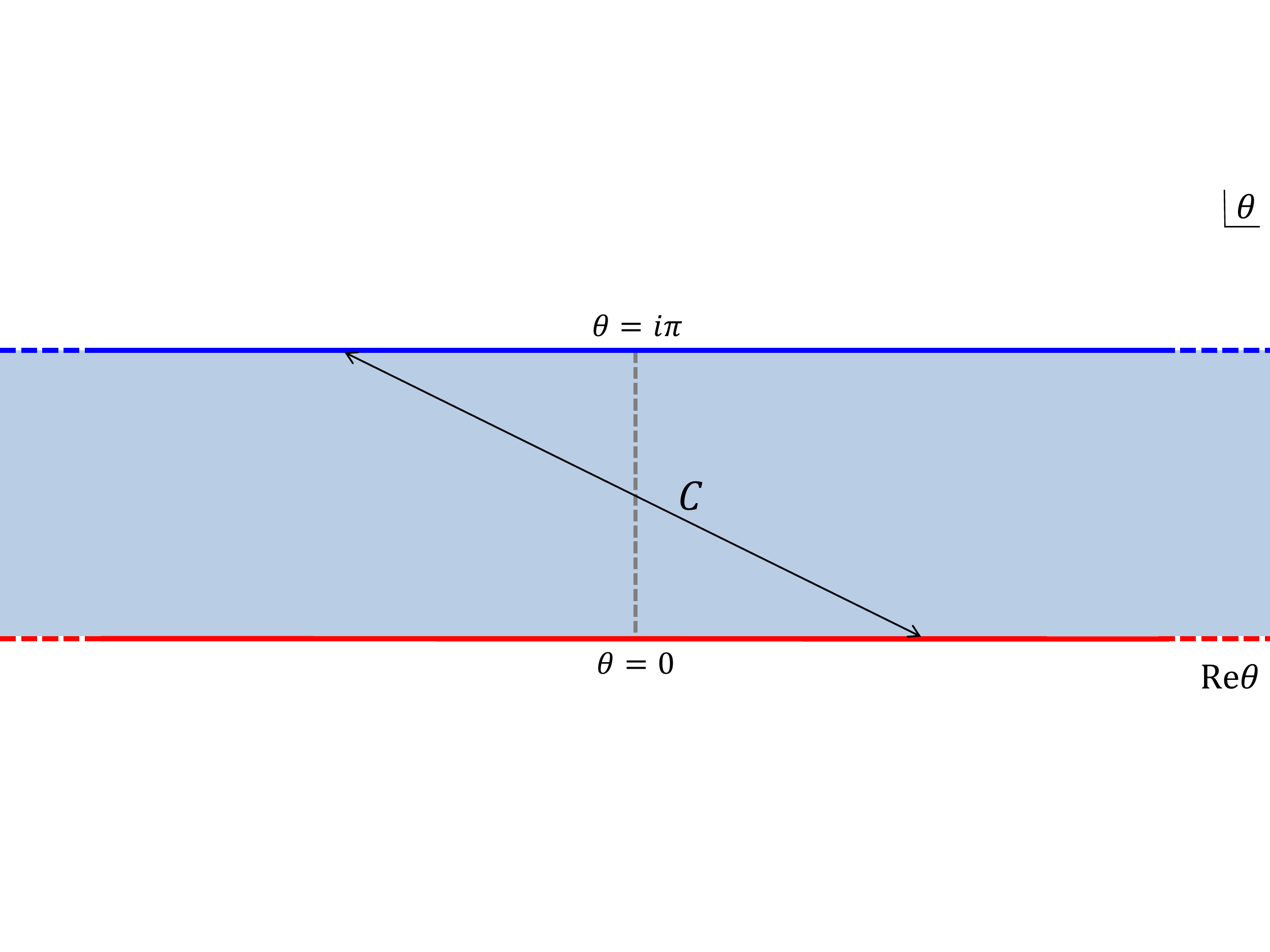}
		\label{thetaplane}
	\end{subfigure}
\begin{subfigure}{.56\textwidth}
	\centering
	\includegraphics[trim=2cm 2cm 2cm 2cm, clip=true, width=\linewidth]{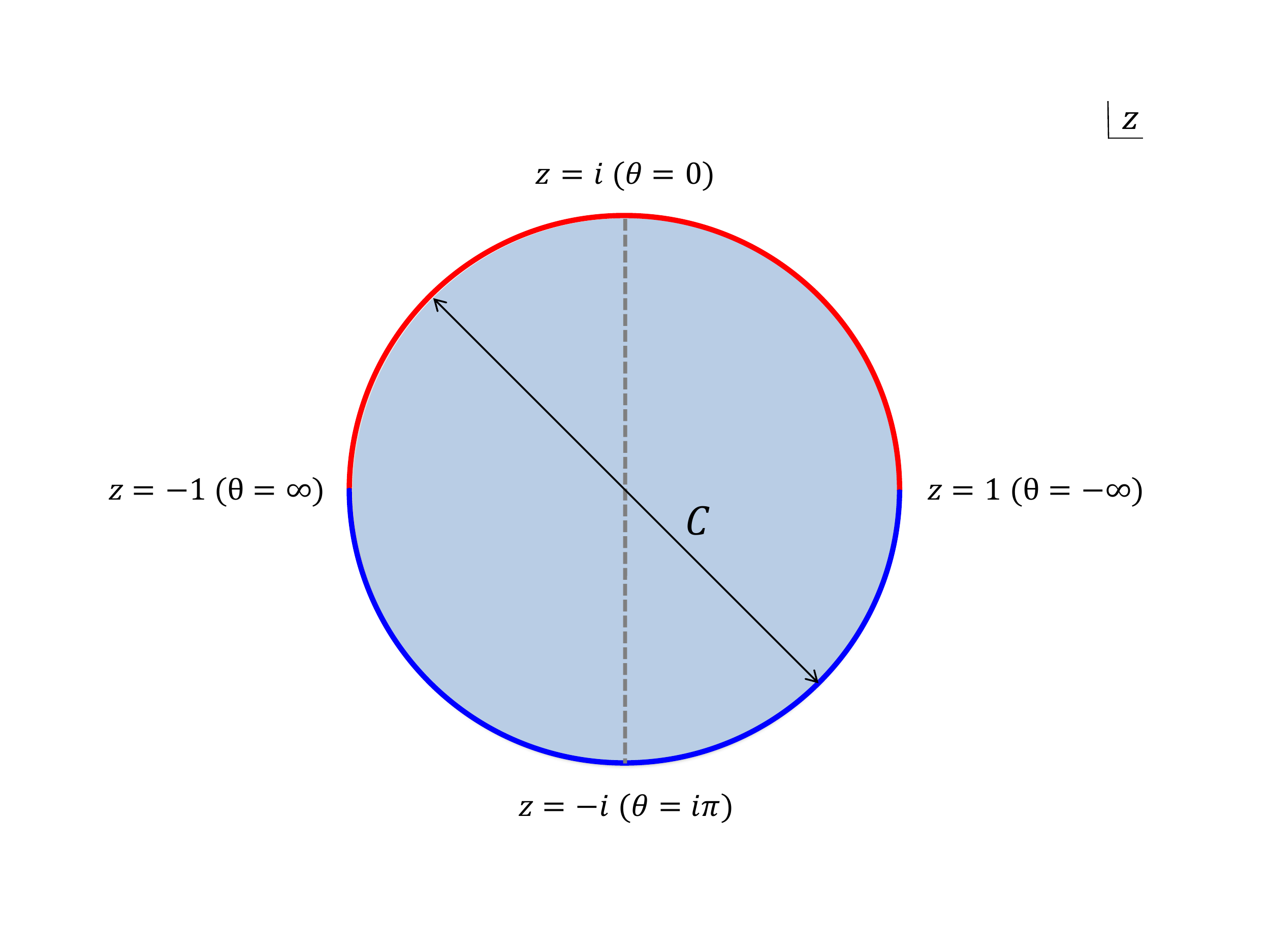}
	\label{zplane}
\end{subfigure}
	\caption{Different variables used to write the $S$-matrices. The first one $s$ is the usual Mandelstam variable and the other two are defined in eq.(\ref{stheta}).}
	\label{sztheta}
\end{figure}

The 2d $O(N)$ non-linear sigma model model was solved by Zamolodchikov and Zamolodchikov in \cite{Zamolodchikov:1978xm}. Its action is given by
\beq
 S = \frac{1}{2g_0^2} \int\!\! d^2x\ (\partial_a \vec{n})^2, \ \ \ \ a=0,1\ ,
\eeq
and $\vec{n}$ is an $N$-dimensional unit vector, $\vec{n}^2=1$. It is an asymptotically free theory that develops a mass gap \cite{ONmodel} and therefore it can be considered as a toy model for QCD. The dynamics is described by $N$ equivalent species of scalar particles with an $O(N)$ symmetry that relates them. There are no bound states, an interesting property for the S-matrix bootstrap maximization program since in previous cases the main idea was to maximize the coupling between particles and their bound states.  In the case of only one spatial dimension, in $2\rightarrow 2$ particle scattering the scattered particles have to have the same momenta as the initial ones. Since we do not initially assume integrability we should allow for possible particle production. 
Consider the $2\rightarrow 2$ scattering matrix for $(p_1,a) + (p_2,b) \rightarrow (p_3,c) + (p_4,d)$, where $p_{1,2}$ ($p_{3,4}$) are the initial (final) momenta and $a,b,c,d$ are $O(N)$ indices. The S-matrix can be written in terms of three functions of $s=(p_1+p_2)^2$, namely $S_T(s)$, $S_R(s)$ and $S_A(s)$ representing the transmission, reflection and annihilation amplitudes:
\begin{equation}
\begin{aligned}
 S_{ab\rightarrow cd}=&\left[ \delta_{ab}\delta_{cd}\ S_A(s)+\delta_{ac}\delta_{bd}\ S_T(s) + \delta_{ad}\delta_{bc}\ S_R(s)\right]   \delta(p_1-p_3) \delta(p_2-p_4)\\
 &+(p_3\leftrightarrow p_4)(c\leftrightarrow d),
 \end{aligned}
 \end{equation}
or equivalently
\begin{equation}
\begin{aligned}
 S_{ab\rightarrow cd}=&
 \left(\frac{1}{N}\delta_{ab}\delta_{cd}S_I(s)+\half(\delta_{ac}\delta_{bd}+ \delta_{bc}\delta_{ad}-\frac{2}{N}\delta_{ab}\delta_{cd}) S_+(s)  \right.\\
 &\left. +\half(\delta_{ac}\delta_{bd}- \delta_{ad}\delta_{bc})S_-(s)\right)  \delta(p_1-p_3) \delta(p_2-p_4)\\
 &+(p_3\leftrightarrow p_4)(c\leftrightarrow d) \ ,
\end{aligned}
\end{equation}
with
\beq\label{atripm}
S_I = N S_A + S_T + S_R, \ \ \ S_\pm = S_T\pm S_R  .
\eeq
 The functions $S_I(s)$ and $S_\pm(s)$ represent the scattering amplitudes in the three isospin channels: isoscalar, symmetric and antisymmetric. The functions $S_a$, $a=I,+,-$ can be analytically continued in the $s$ plane, defining the physical sheet. They have a cut for real $s$ and $s>4m^2$ (the physical line) and another cut for real $s$ and $s<0$ corresponding to the $t$-channel (in 2d, $t=4m^2-s$). The values of $S_a$ right above and below the cut are related 
 by the real analyticity constraint:
 \beq
  S_a(s+i 0^+) =\left[ S_a(s-i 0^-) \right]^*, \ \ \ \ s\in \mathbb{R},\ \  s>4m^2.
 \eeq 
 It is convenient to use the relative rapidity variable $\theta$ defined through
\beq
 s = 4 m^2 \cosh^2\frac{\theta}{2},  \ \ \  z=\frac{i-e^\theta}{i+e^\theta},
 \label{stheta}
\eeq
where we also introduced the variable $z$ since it will be useful later (see Figure \ref{sztheta}). The physical line is then $\theta\in \mathbb{R}$, or $z=e^{i\phi}, 0\le\phi\le\pi$ and the physical region $0\le \Im\theta\le\pi$, or $|z|\le1$. The crossing symmetry relates the functions at $\theta$ and $i\pi-\theta$ (or $z$ and $-z$) as $S_T(i\pi-\theta)=S_T(\theta)$,  $S_R(i\pi-\theta)=S_A(\theta)$. In terms of the isospin channels the crossing symmetry reads
\beq
 S_a(i\pi-\theta) = \sum_b C_{ab} S_b(\theta), \ \ \ \ \ a,b = I,+,- ,
\eeq
where from now on the indices $a,b,\ldots$ label isospin channels and the matrix $C$ satisfies $C^2=1$ and is given by
\beq\label{ipmcrossing}
C = 
 \left(\begin{array}{ccc} \frac{1}{N} & \frac{N}{2}+\half-\frac{1}{N} &\half-\frac{N}{2}\\ 
 	                                  \frac{1}{N} & \half -\frac{1}{N}                   & \half  \\ 
 	                                 -\frac{1}{N} & \half+\frac{1}{N}                   & \half  \end{array}\right)  .
\eeq
 It is natural to diagonalize $C$ by introducing the functions
\beqa
S_1 &=& S_T = \half (S_+ + S_-)\ , \\
S_2 &=&  \half (S_R + S_A) = \frac{1}{2N} S_I +\frac{N-2}{4N} S_+ -\frac{1}{4} S_- \ , \\
S_3 &=& \half (S_R - S_A) = - \frac{1}{2N} S_I+\frac{N+2}{4N} S_+ -\frac{1}{4}S_- \ ,
\label{S123}
\eeqa
such that $S_{1,2}(i\pi-\theta)=S_{1,2}(\theta)$, $S_{3}(i\pi-\theta)=-S_{3}(\theta)$. Up to now we have only described general constraints on the S-matrix due to standard properties of the field theory.  

\subsection{Exact solution using the Yang-Baxter equation}

Now let us describe the exact S-matrix obtained in \cite{Zamolodchikov:1978xm} by using the Yang-Baxter equation. It can be written in terms of the function 
\beq
 Q(\theta) = \frac{\Gamma\left(\frac{\lambda-i\theta}{2\pi}\right)\Gamma\left(\half-\frac{i \theta}{2\pi}\right)}{\Gamma\left(\half+\frac{\lambda-i\theta}{2\pi}\right)\Gamma\left(-\frac{i \theta}{2\pi}\right) }\ ,
\eeq
with
\beq
 \lambda = \frac{2\pi}{N-2}\ .
\eeq
The exact S-matrix reads
\beqa
 S_T(\theta)  &=& Q(\theta) Q(i\pi-\theta) \ ,\\
 S_A(\theta) &=& -\frac{i\lambda }{i\pi-\theta} S_T(\theta) \ , \\
 S_R(\theta) &=&  -\frac{i\lambda }{\theta} S_T(\theta)  \ .
\eeqa
 From here one can compute $S_{I,+,-}$ using eq. \eqref{atripm}. In the physical region $0\le\Im\,\theta\le \pi$, $S_+$ has a zero at $\theta=i\lambda$, both $S_\pm$ have a zero at $\theta=i\pi$ and none of them has poles. All other zeros and any poles are outside the physical region. Therefore, in this case, the``minimal" solution, namely the one without poles in the physical region, corresponds directly to the $O(N)$ sigma model. Multiplying by a CDD factor \cite{Castillejo:1955ed} one can obtain the S-matrix of the Gross-Neveu model \cite{Zamolodchikov:1978xm}. In the rest of this paper we analyze the general constraints on the S-matrix and how the minimal solution, namely the $O(N)$ sigma model can be reproduced by a maximization procedure without any input from integrability. 

\section{The convex space of allowed S-matrices}

 Given a general $S$-matrix the unitarity constraint is
\beq
 S S^\dagger = \mathbb{I}\ .
\eeq
 Consider a subsector $D$ of the space of states and a given state $\ket{\psi_{\alpha}}\in D$ then we have
\beq
 \sum_{\beta \in D} \bra{\psi_\alpha} S \ket{\psi_\beta} \bra{\psi_\beta} S^\dagger \ket{\psi_\alpha}  - \bra{\psi_\alpha}\psi_\alpha\rangle =  - \sum_{\beta \notin D} \left|\bra{\psi_\alpha} S \ket{\psi_\beta} \right|^2 \le 0 \ .
\eeq
Namely, the matrix $S_D$ defined as $S$ projected to the subspace $D$ satisfies
\beq
 S_D S_D^\dagger \preceq \mathbb{I}\ .
\eeq
Consequently, the difference $S_D S^\dagger_D - \mathbb{I}$ is negative definite. This is equivalent to the positive semi-definite condition
\beq
\left(\begin{array}{cc} \mathbb{I} & S_D \\ S_D^\dagger & \mathbb{I} \end{array} \right) \succeq 0 \ .
\eeq
Since the space of positive semi--definite matrices is a convex space, so is the space of allowed S-matrices. In the case discussed in this paper, the subsector $D$ will correspond to two-particle states.  Using various trial wave-functions, it is easy to see that these constraints read simply 
 \beq
 |S_I(\theta)|^2\le 1, \ \ \ \ |S_+(\theta)|^2\le 1,  \ \ \ \ |S_-(\theta)|^2\le 1,  \ \ \ \theta\in \mathbb{R}.\
 \eeq
 Intuitively, for any given total isospin state and fix energy and center of mass momentum ($P_{CM}=0$), there is only one two-particle state. Therefore $|S_a(\theta)|^2\le 1$, $\theta\in\mathbb{R}$ since that channel cannot receive contributions from any initial two-particle state other than itself.  
Thus, if we parameterize the space of allowed S-matrices by the (complex) values $S_a(\theta=\sigma), \sigma\in\mathbb{R}$, namely on the physical line, then each variable is restricted to the unit disk which is a convex space. The space is further restricted by the linear crossing constraint \eqref{ipmcrossing}. Therefore, the space can be described as the intersection of the (infinite) product of unit disks and the linear subspace defined by crossing. In the following section we characterize this space more precisely. 

\section{Maximization of a linear functional in a convex space}

 Let us start with a simple example of convex maximization. Consider a convex polyhedron in $\mathbb{R}^n$ defined by linear constraints that are given by normals $v^{j}_A$: $\sum_{A=1}^n v^{j}_A x_A\le b^j$, where $x_A \in \mathbb{R}^n$ and $b^j$ are real constants. Any generic linear function $F_w=\sum_{A=1}^n w_A x_A$ inside such polyhedron attains its maximum at a unique vertex on the boundary (for exceptional values of $w_A$ the maximum can be an edge or a face). This property of convex maximization allows for efficient numerical algorithms to obtain the maximum, in particular in this work we used \cite{cvx}.
 
   In this section we argue that the problem at hand is a generalization of this simple problem to infinite dimensions and quadratic constraints. The analogy is made clearer by using a notation where the index $A$ should be thought as $(a,\sigma)$ or $(a,j)$. Here $a=I,+,-$ labels the different functions in the S-matrix and $\sigma$ is a continuous index that labels the physical line. Alternatively, as described in the appendix, the index $j=1\ldots M$ labels a set of points on the physical line used to interpolate the functions and obtain a discretized version of the problem, \ie\ $M\rightarrow\infty$  recovers the continuum. Also, we do not use the repeated index convention, sums over $A$ are  written explicitly and in the case of $\sigma$ should be understood as integrals. 
 
 With this in mind, the unitarity constraints are 
 \beq
  \ \ |S_A|^2 = (\Re S_A)^2 +  (\Im S_A)^2 \le 1\ ,
 \eeq
 for every value of the index $A$.  For any analytic function in a region, the real part at the boundary is enough to determine the function inside and, in particular, the imaginary part at the boundary. Thus, if $\Re S_A$, is given, namely the real part of the S-matrices on the real line, using crossing symmetry $S_a(i\pi-\sigma)=\sum_b C_{ab}S_b(\sigma)$ the real part on the line $\theta=i\pi-\sigma$ can be computed. Since now the real part is known in the whole boundary of the physical region one can compute the imaginary part. This results in a linear relation 
 \beq
  \Im S_A = \sum_B  K_{AB} \Re S_B
 \eeq
 for some kernel $K_{AB}$. In the $\theta$ plane, the explicit form of the kernel $K_{AB}$ is
\beqa
 \Im S_a(\sigma) &=& -\frac{1}{2\pi} \sum_b \fint_{-\infty}^{\infty} \left[ \frac{\cosh(\frac{\sigma'+\sigma}{2})}{\sinh(\frac{\sigma'-\sigma}{2})} \delta_{ab}
- \frac{\sinh(\frac{\sigma'-\sigma}{2})}{\cosh(\frac{\sigma'+\sigma}{2})} C_{ab} \right]\Re S_b(\sigma') \frac{d\sigma'}{\cosh\sigma'} \non \\
&& + \Im S_a(\theta=\frac{i\pi}{2}) \\
 &=& -\frac{1}{2\pi} \sum_b \int_{-\infty}^{\infty} \left(\delta_{ab} \coth(\frac{\sigma'-\sigma}{2}) + C_{ab} \tanh(\frac{\sigma'+\sigma}{2}) \right) \Re S_b(\sigma') \non \\
 & &+\frac{1}{2\pi}\sum_b \int_{-\infty}^{\infty} \left(\delta_{ab} +C_{ab}\right) \Re S_b(\sigma') \tanh \sigma' + \Im S_a(\theta=\frac{i\pi}{2})\ .
 \label{Kernel}
\eeqa
In both formulas we included the term $\Im S_a(\theta=\frac{i\pi}{2})$ for completeness. In the case of interest here $\Im S_a(\theta=\frac{i\pi}{2})=0$ by real analyticity; therefore, we ignore this term in the discretized version used for numerical computations in the appendix. Now, the unitarity constraints read
\beq
 |S_A|^2 = \left(\Re S_A\right)^2 + \left (\sum_B  K_{AB} \Re S_B\right)^2 = \sum_{BC} \Re S_B\,  H^A_{BC}\, \Re S_C \le 1\ ,
\eeq
where
\beq
 H^A_{BC} = \delta^{A}_{B} \delta^{A}_{C} + K_{AB} K_{AC}\ .
\eeq
 Since $|S_A|^2\ge 0$, these are positive semi-definite quadratic forms labeled by the index $A$.  Therefore the intersection is a convex space. It is non-empty since the origin $\Re S_A=0$ is in the space, and it is contained in the hypercube $|\Re S_A|\le1$. The (curved) faces of this generalized polyhedron are given by the points that satisfy all constraints but saturate only one (namely for one value of $A$). The edges are determined by the points saturating all constraints except for one value of $A$.  

\begin{figure}
 	\centering
 	\includegraphics[width=0.5\textwidth]{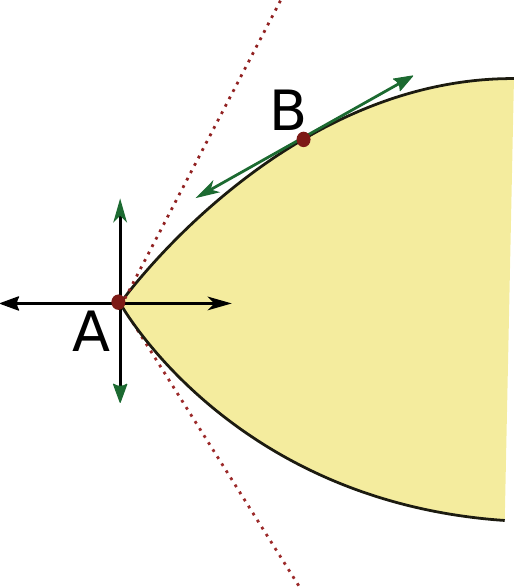}
 	\caption{Schematic representation of the convex space of allowed S-matrices. The non-linear sigma model is at a vertex like $A$ where there are no directions such that at the linearized level satisfy the constraints in both directions. At a point like $B$ there are tangents that are parallel to the boundary and appear as zero modes of the linearized constraints. The dotted lines represent the linearized constraints described by $V^C_{B} \xi_B \le 0$.}
 	\label{Vdef}
 \end{figure}

\begin{figure}
\includegraphics[trim=0cm 0cm 0cm 0cm, clip=true, width=\textwidth]{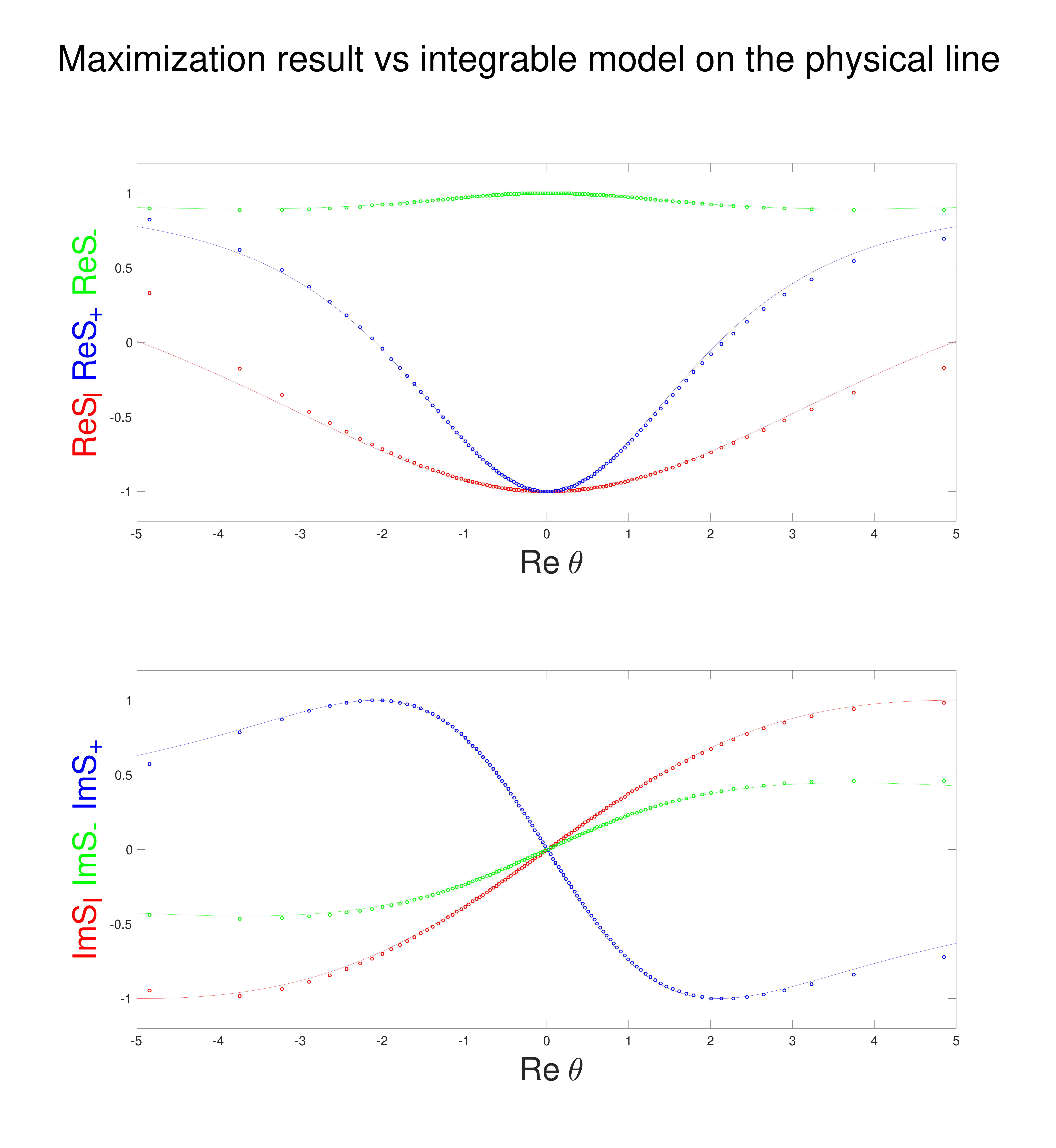}
\caption{The comparison of the maximization results with the exact integrable model for $N=4$. We use $200$ points for interpolation and choose an initial functional $F=\Re[-S_1(\theta_0)-\alpha S_2(\theta_0)]$ with $\theta_0\simeq i(z_0=0.3i)$ and $\alpha=100$. Here we plot the real and imaginary parts of $S_I$, $S_+$ and $S_-$ on the physical line where the dots indicate maximization results while the curves are the exact integrable model.}
\label{N=4}
\end{figure}

\begin{figure}
\includegraphics[trim=0cm 0cm 0cm 0cm, clip=true, width=\textwidth]{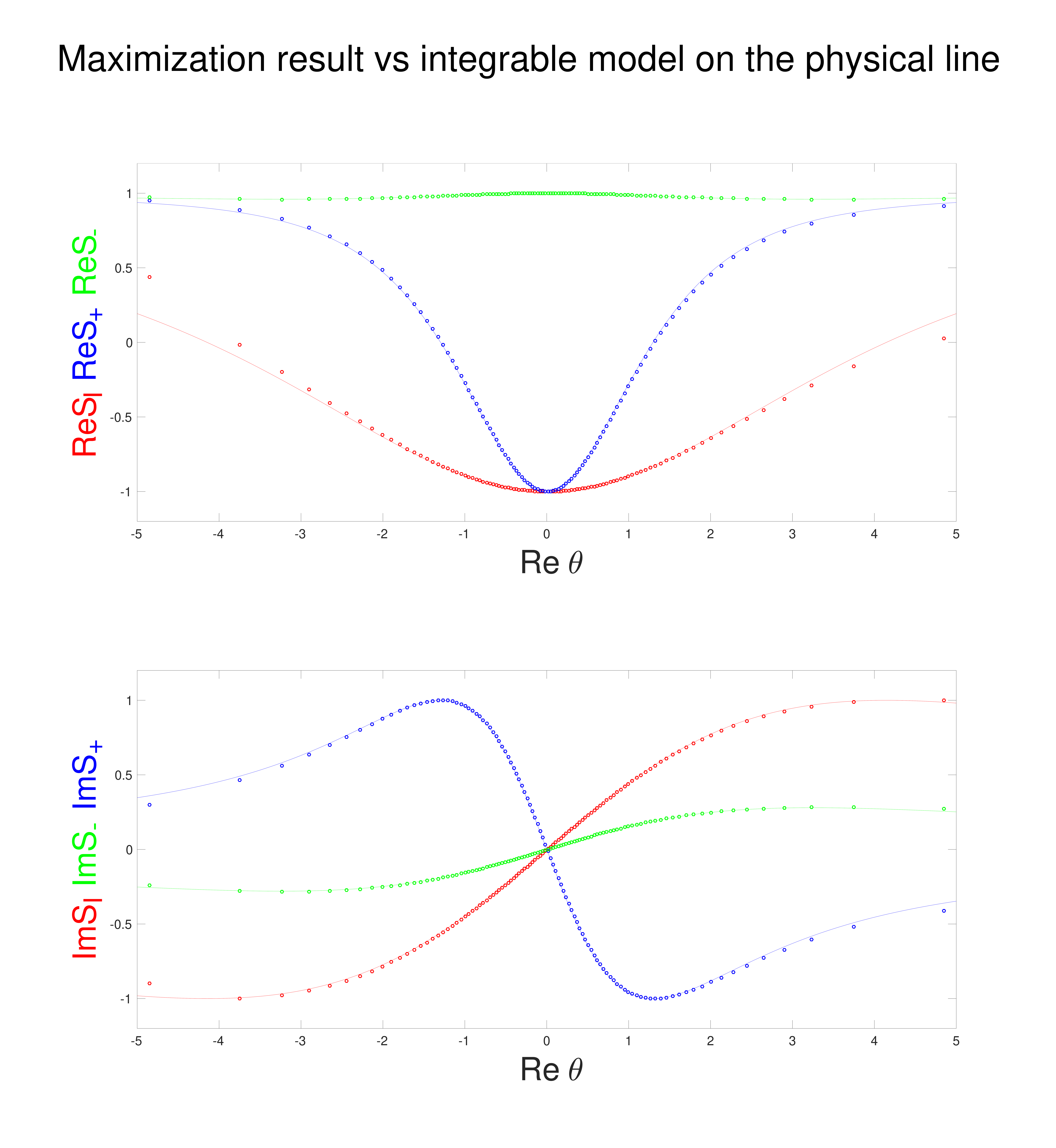}
\caption{The comparison of the maximization results with the exact integrable model for $N=6$. We use $200$ points for interpolation and choose an initial functional $F=\Re[S_1(\theta_0)-\alpha S_2(\theta_0)]$ with $\theta_0\simeq i(z_0=0.3i)$ and $\alpha=7.5$. Here we plot the real and imaginary parts of $S_I$, $S_+$ and $S_-$ on the physical line where the dots indicate maximization results while the curves are the exact integrable model.}
\label{N=6}
\end{figure}

\begin{figure}
\includegraphics[trim=0cm 0cm 0cm 0cm, clip=true, width=\textwidth]{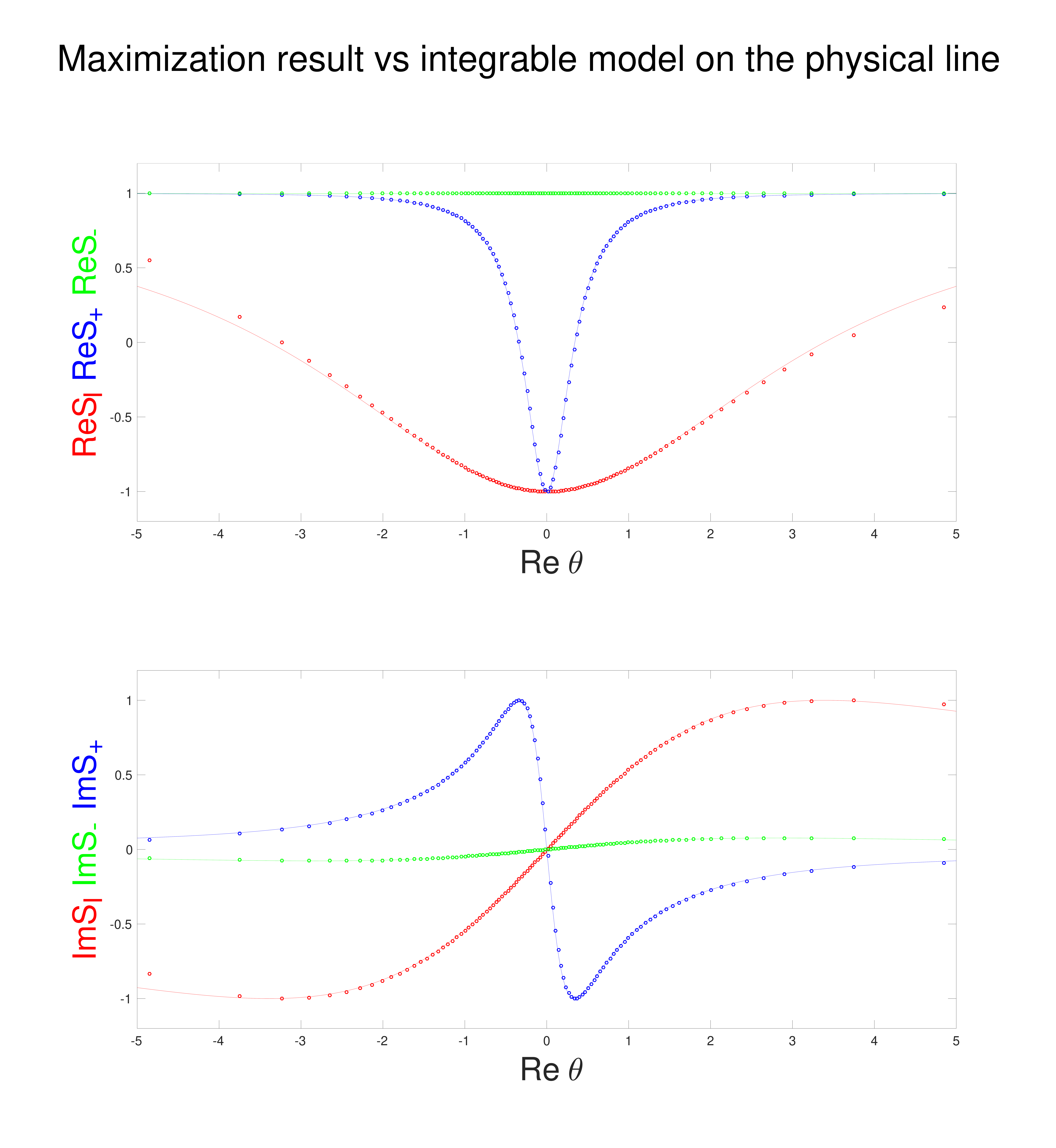}
\caption{The comparison of the maximization results with the exact integrable model for $N=20$. We use $200$ points for interpolation and choose an initial functional $F=\Re[S_1(\theta_0)-\alpha S_2(\theta_0)]$ with $\theta_0\simeq i(z_0=0.3i)$ and $\alpha=3$. Here we plot the real and imaginary parts of $S_I$, $S_+$ and $S_-$ on the physical line where the dots indicate maximization results while the curves are the exact integrable model.}
\label{N=20}
\end{figure}

\begin{figure}
\includegraphics[trim=0cm 0cm 0cm 0cm, clip=true, width=\textwidth]{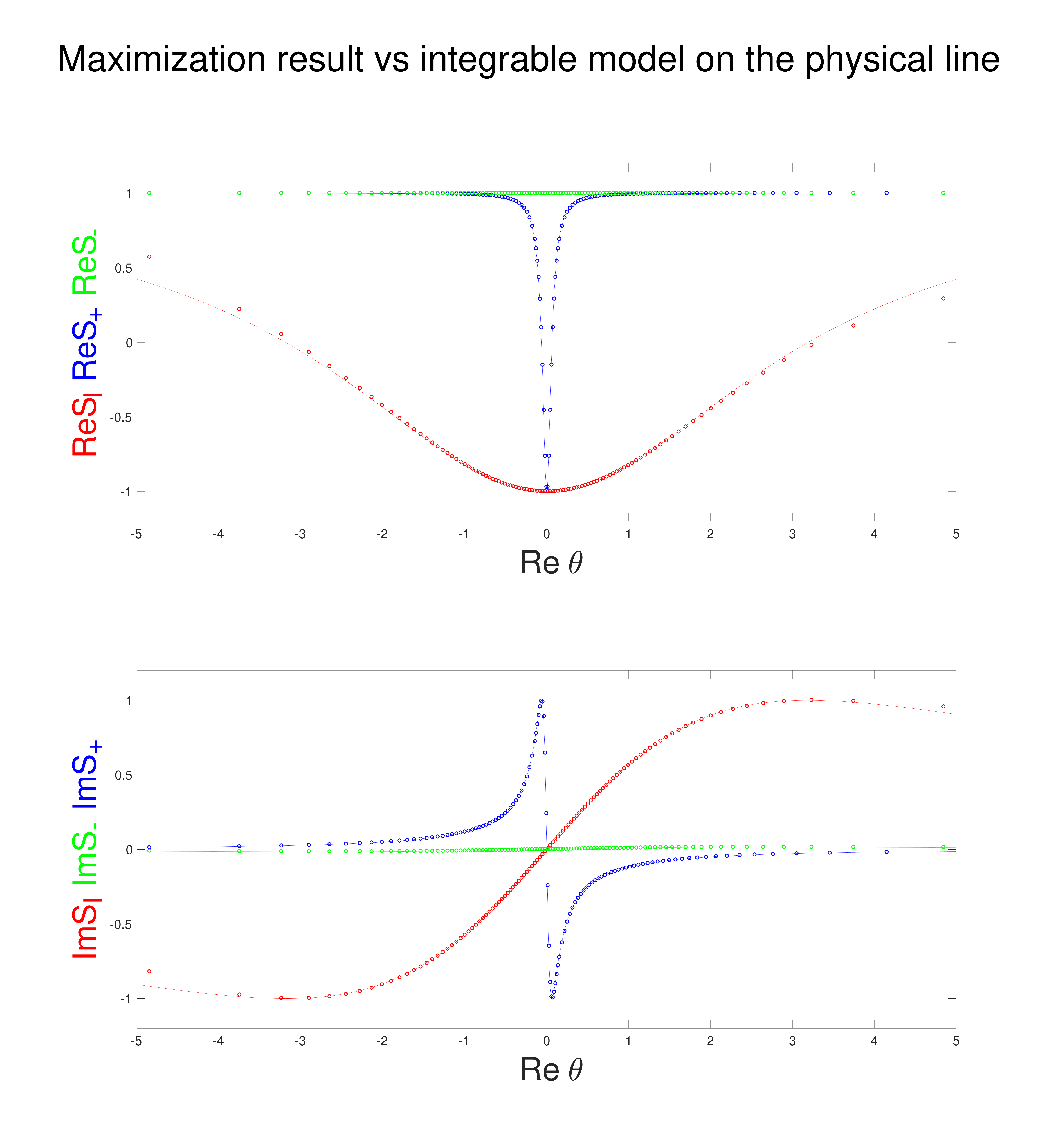}
\caption{The comparison of the maximization results with the exact integrable model for $N=100$. We use $200$ points for interpolation and choose an initial functional $F=\Re[S_1(\theta_0)-\alpha S_2(\theta_0)]$ with $\theta_0\simeq i(z_0=0.3i)$ and $\alpha=2.4$. Here we plot the real and imaginary parts of $S_I$, $S_+$ and $S_-$ on the physical line where the dots indicate maximization results while the curves are the exact integrable model.}
\label{N=100}
\end{figure}

The integrable model saturates all the constraints and therefore is at the boundary of this space. Moreover, since the integrable model has no free parameters, we do not expect to have any other near-by point that saturates all constraints and therefore the integrable model should be at a vertex (or corner) of this space. In fact this expectation can be checked numerically. By the discussion at the beginning of the section, to find such a vertex, it is sufficient to maximize a linear functional
\beq\label{maxfunc}
 F_w = \sum_A w_A \Re S_A\ ,
\eeq
such that $w^A$ points in the general direction of the vertex. Near the vertex that we take to be at $\Re S_A = \Re\hat{S}_A$ the space has the shape of a cone that we can obtain by linearizing the quadratic forms using a perturbation $\xi_A$ around the maximum:
\beq
 \Re S_A = \Re \hat{S}_A + \xi_A \ .
\eeq
 It follows that the space of S-matrices is given, at the linearized level (see fig.\ref{Vdef}), by the constraints 
\beq
  V^C_{B} \xi_B \le 0, \ \ V^C_{B} = \sum_A \Re \hat{S}_A H^{C}_{AB},\ \forall C.
\eeq
 The matrix $V^C_B$ can be thought as a set of vectors labeled by $C$ and normal to the faces of the cone whose tip is at the vertex.
 The edges are given by allowing only one constraint not to be saturated. Arranging the vectors parallel to the edges in a matrix we get the inverse matrix of $V$. Numerically, once a reasonable direction is found it can be improved iteratively by taking
 \beq\label{w}
  w_A = \sum_B V^B_{A}\ .
 \eeq
 The rationale is that such direction is the sum of the normals to the faces and therefore should point in the direction of the vertex. The numerical calculations agree well with this naive expectation. 
There are many choices for the initial direction $w_A$ but a little experimentation shows that adequate functionals are
\beq
 F = \Re [ S_1(\theta_0) - \alpha S_2(\theta_0) ]\ ,
\eeq
Here, $S_{1,2}$ are the functions defined in eq.(\ref{S123}). There are various choices for the point $\theta_0$. One we choose frequently is $\theta_0\simeq i$ (corresponding to  $z_0 = 0.3 i$) but it does not have to be on the imaginary axis.  
\begin{table}[h]
	\centering
	\renewcommand{\arraystretch}{1.2}
	\begin{tabular}{@{}rrc@{}}
		\toprule
	     N & $\alpha$ & \\ 
		\midrule
           5 &     13 & \  \\
	       6 &\ \ 7.5    & \\
	       7 & 6    &\\
	       8 & 5   &\\
	     20 & 3  &\\
	   100 & 2.4 &\\
		\bottomrule
	\end{tabular}
	\caption{Values for $\alpha$ coefficient in maximization function}
	\label{alphaCoeff}
\end{table} 
There is a range of coefficients $\alpha$ that can be chosen depending on $\theta_0$ and $N$. We give  suggested values in Table \ref{alphaCoeff}. As mentioned, the functional, namely the choice of vector $w_A$, can be improved iteratively using eq.(\ref{w}).
After a few iterations this leads to a rapid decrease of the error as measured by how close the solution is to saturate the unitarity bounds. After the iterative procedure converges, the maximum of the functional is simply
\beq
 F^{\rm it}_{\rm max} = 3M
\eeq
where $M$ is the number of interpolating points on the physical line. The reason is that the integrable model $\hat{S}_A$ saturates all the constraints and therefore
\beq
 F^{\rm it}_{\rm max} = \sum_A w_A \Re \hat{S}_A = \sum_{AB} V^B_A \Re \hat{S}_A = \sum_{ABC} \Re\hat{S}_C H^{B}_{CA} \Re \hat{S}_A =\sum_B 1 = 3M.
\eeq
Finally, we can compare the maximization results with the exact integrable model as shown in Figures \ref{N=4}-\ref{N=100}. It is clear that there is a very good agreement.
 We emphasize again that this agreement was obtained by maximizing a functional without making any assumptions on the S-matrices such as position of its  zeros. Only the constraints of unitarity, crossing and real analyticity were used. 

Finally, for the integrable model to be at a vertex, the matrix $V^B_A$ should have no zero modes. The reason is that we cannot move away from a vertex and still saturate all constraints (see Figure \ref{Vdef}). If there is a zero mode $\xi^{(0)}_A$ such that
\beq
 V^B_A \xi^{(0)}_A = 0
\eeq 
then a deformation along $\xi^{(0)}_A$ will still saturate all constraints. Numerically we checked that the matrix $\tilde{V}=V^T V$ has no zero eigenvalues in agreement with $V$ having no zero modes.

\begin{figure}
\includegraphics[trim=0cm 5cm 0cm 5cm, clip=true, width=\textwidth]{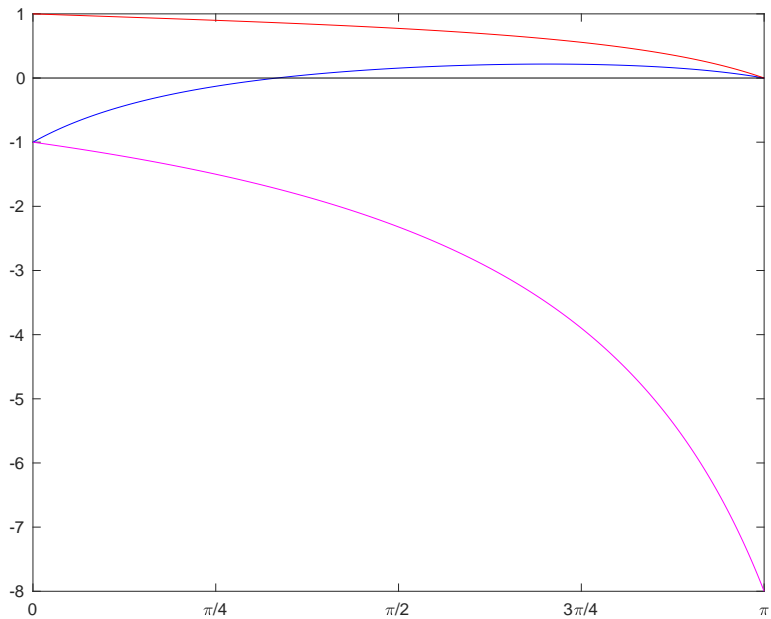}
\caption{Plot of the maximization results on the imaginary axis of $\theta$ for $N=8$. It is easy to see that $S_-$ (red) has a zero on the imaginary axis at $\theta\simeq i\pi$ and $S_+$ (blue) has two zeros, one at $\theta\simeq i\pi$ and the other at $\theta\simeq1.048 i$. Finally, $S_I$ (purple) has no zeros.}
\label{imag}
\end{figure}

\begin{figure}
	\includegraphics[trim=0cm 5cm 0cm 5cm, clip=true, width=\textwidth]{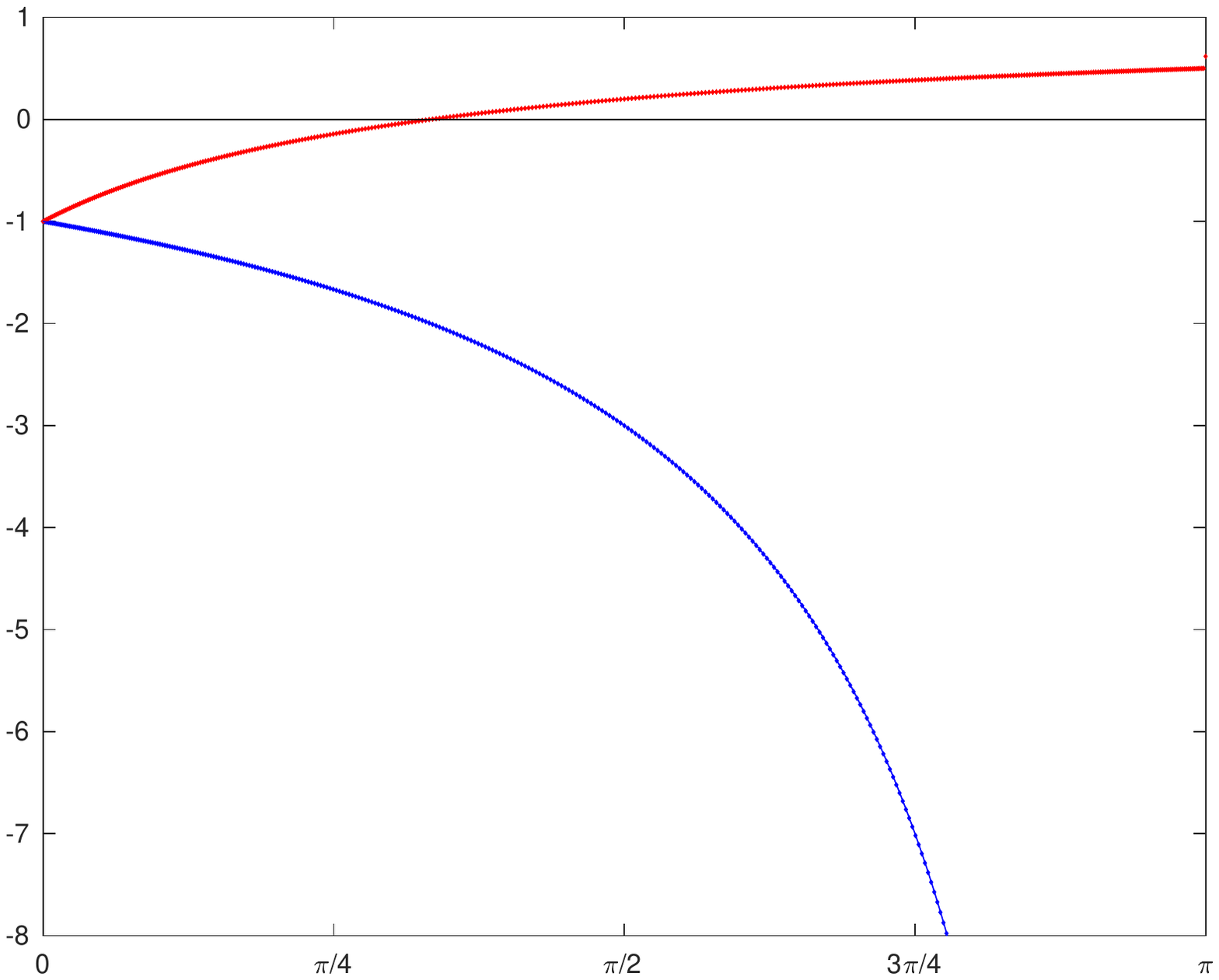}
	\caption{Plot of the numerical ratios $R_1=\frac{S_I}{S_-}$ (blue) and $R_2=\frac{S_+}{S_-}$ (red) on the imaginary axis together with the fit based on the naive ratios obtained from the zeros in the previous figure: $\tilde{R}_1=\frac{\theta+i\pi}{\theta-i\pi}$, $\tilde{R}_2=\frac{\theta-1.048 i}{\theta+1.048 i}$. The numerical results and the fits are indistinguishable.}
	\label{imagfit}
\end{figure}

\subsection{Analytic results from the numerics}
 
 In this section we show how one could extend the numerical solution into an analytical one by using some simple inputs from the numerics. This might seem as a moot point since the analytical solution is already known but it is useful to see how the solution arises without using factorization. 
 The numerical solution clearly shows that there is a zero of $S_+$ in the imaginary axis of $\theta$ and that $S_\pm$ are zero at (or near) $\theta=i\pi$. See Figure \ref{imag}. Moreover, since the functions are related by crossing, it is clear that they should have most zeros and poles in common. We then expect that the ratios 
 \beq
  R_1 = \frac{S_I(\theta)}{S_-(\theta)}, \ \ \ R_2 = \frac{S_+(\theta)}{S_-(\theta)}\ ,
 \eeq
have a small number of zeros and poles. Actually since the $S$-matrices saturate the constraints ($|S_a(\theta)|=1$, $\theta\in\mathbb{R}$), then $|R_1(\theta)|=|R_2(\theta)|=1$ on the real axis, the zeros and poles should be situated symmetrically with respect to the real axis:
\beq
 R_1 = \prod_{\ell=1}^{n_1} \frac{\theta-\theta_\ell}{\theta-\theta^*_\ell}, \ \ \ \ \ 
 R_2 = \prod_{\ell=1}^{n_2} \frac{\theta-\tilde{\theta}_\ell}{\theta-\tilde{\theta}^*_\ell}\ .
\eeq
 As discussed, the numerical solution is such that, on the imaginary axis, $S_I$ has no zeros, $S_-$ has one zero at $i\pi$ and $S_+$ has two zeros, one at $i\pi$ and the other at an intermediate point 
 $\theta=i \tau_0$ that can be determined from the plot (\eg\ Figure \ref{imag} for $N=8$). The naive ratios that take these zeros into account fit the numerical results perfectly well 
\beq
 R_1 = \frac{\theta+i\pi}{\theta-i\pi}, \ \ \ \ R_2 =  \frac{\theta-i\tau_0}{\theta+i\tau_0}\ ,
\eeq
as seen in Figure \ref{imagfit}. From the crossing conditions:
\beqa
R_1(i\pi-\theta)&=&\frac{\frac{1}{N} R_1(\theta)+(\frac{N}{2}+\half-\frac{1}{N}) R_2(\theta)+\frac{1-N}{2}}{-\frac{1}{N} R_1(\theta)+(\half+\frac{1}{N}) R_2(\theta)+\half}\\
R_2(i\pi-\theta)&=&\frac{\frac{1}{N}R_1(\theta)+(\half-\frac{1}{N})R_2(\theta)+\half}{-\frac{1}{N}R_1(\theta)+(\half+\frac{1}{N})R_2(\theta)+\half}\ ,
\eeqa
  it results that the only solution where $R_{1,2}$ are ratios of linear functions is the one described with one zero at $i\pi$, and the other at exactly $\tau_0=\lambda=\frac{2\pi}{N-2}$, in good agreement with the numerics. It is now convenient to use as a reference the crossing invariant function $S_2$ in eq.(\ref{S123}). Some simple algebra gives 
 \beqa
  S_I  &=& -\frac{N-2}{\pi^2} (\theta+i\pi)(\theta+i\lambda)\, S_2, \\
  S_+ &=& -\frac{N-2}{\pi^2} (\theta-i\pi)(\theta-i\lambda)\,  S_2, \\
  S_-  &=& -\frac{N-2}{\pi^2} (\theta-i\pi)(\theta+i\lambda)\, S_2.
\eeqa
Since $S_2(i\pi-\theta)=S_2(\theta)$ and the poles and zeros of $S_{I,+,-}$ are symmetric with respect to the real axis, it is clear that, for example, in $S_+$, the zero at $i\lambda$ implies a pole at $-i\lambda$ that should be in $S_2$. But then $S_2$ should have a pole at $i\pi+i\lambda$ that appears in $S_+$ implying a zero at $-i\pi-i\lambda$ and so on. Thus, the factor $(\theta-i\lambda)$ should be extended as   
\beqa
 (\theta-i\lambda) &\rightarrow& \frac{ (\theta-i\lambda)}{(\theta+i\lambda)} \frac{ (\theta+i\pi+i\lambda)}{(\theta-i\pi-i\lambda)}
 \frac{ (\theta-2i\pi-i\lambda)}{(\theta+2i\pi+i\lambda)} \frac{ (\theta+3i\pi+i\lambda)}{(\theta-3i\pi-i\lambda)}  \ldots \\
 && = \frac{\Gamma\left(-\frac{i\theta}{2\pi}+\frac{\lambda}{2\pi}\right)\Gamma\left(\frac{i\theta}{2\pi}+\frac{\lambda}{2\pi}+\half\right)}{\Gamma\left(-\frac{i\theta}{2\pi}+\frac{\lambda}{2\pi}+\half\right)\Gamma\left(\frac{i\theta}{2\pi}+\frac{\lambda}{2\pi}\right)}\ ,
\eeqa
and similarly with the factor $(\theta-i\pi)$. If we were to do the same in $S_I$ the factor $(\theta+i\lambda)$ should be extended without generating a pole at $i\lambda$ and that is accomplished by having the pole at $-i\lambda$, namely the same $S_2$ as needed in $S_+$. Finally, by normalization, the overall factor $\frac{N-2}{\pi^2}$ should be absorbed in $S_2$. All in all we get for $S_+$
\beq
 S_+ = - \frac{\Gamma\left(-\frac{i\theta}{2\pi}+\frac{\lambda}{2\pi}\right)\Gamma\left(\frac{i\theta}{2\pi}+\frac{\lambda}{2\pi}+\half\right)}{\Gamma\left(-\frac{i\theta}{2\pi}+\frac{\lambda}{2\pi}+\half\right)\Gamma\left(\frac{i\theta}{2\pi}+\frac{\lambda}{2\pi}\right)}
 \frac{\Gamma\left(-\frac{i\theta}{2\pi}+\half\right)\Gamma\left(\frac{i\theta}{2\pi}+1\right)}{\Gamma\left(-\frac{i\theta}{2\pi}+1\right)\Gamma\left(\frac{i\theta}{2\pi}+\half\right)}\ ,
\eeq 
from where we can compute $S_I$ and $S_-$  satisfying all the required properties and in agreement with the integrable model. It seems that one can find other functions that saturate all constraints by assuming that $R_{1,2}$ are ratios of higher order polynomials since that appears to allow more freedom given that we have more roots to choose. However it is easy to see that the number of equations grows faster suggesting that it is not easy to solve the crossing constraints in terms of other than linear polynomials. 

\section{Saturating the unitarity constraints}

As we saw, the integrable model saturates unitarity, however they are not the only functions to do that. In this subsection we discuss other functions that also saturate those constraints as a way to study the boundary of the space of allowed S-matrices.

\subsection{The CDD factors}\label{CDD}

The 2d $O(N)$ bosonic integrable model admits a CDD type ambiguity \cite{Castillejo:1955ed,Zamolodchikov:1978xm}. The minimal solution can be multiplied by any number of CDD factors:
\begin{equation}
f(\theta,\alpha)=\frac{\sinh\theta+i\sin\alpha}{\sinh\theta-i\sin\alpha}
\end{equation}
and the result preserves crossing symmetry and unitarity saturation. Notice that a CDD factor will introduce a pair of poles into the physical strip at $\theta=i\alpha$ and $\theta=i\pi-i\alpha$ if $\alpha$ is taken to be real and $\Re\,\alpha\in[0,\pi]$. In particular one can use such a factor to obtain the Gross-Neveu model S-matrices \cite{Zamolodchikov:1978xm}.  Here we want to stay in the space of functions analytic in the physical region and avoid adding any poles. One can then construct S-matrices of the type
\begin{eqnarray}
\tilde{S}_I(\theta)=\prod_i f(\theta,\alpha_i) S_I(\theta),\\
\tilde{S}_+(\theta)=\prod_i f(\theta,\alpha_i) S_+(\theta),\\
\tilde{S}_-(\theta)=\prod_i f(\theta,\alpha_i) S_-(\theta).
\end{eqnarray}
 These are at the boundary of the convex space because they satisfy all the constraints and saturate the unitarity condition. We use the explicit form of such CDD dressed integrable model to construct the $w_A$ according to eq.\eqref{w} and indeed we find them numerically through maximizing the functional eq.\eqref{maxfunc}. We also check the rigidity of the CDD dressed integrable model in this space, namely, whether it is a vertex. We find that the matrix
\begin{equation}
\tilde{V}=V^{T}V
\label{VTV}
\end{equation}
has $6$ zero modes for each CDD factor which corresponding to $6$ independent infinitesimal deformation that preserve the crossing and unitarity condition. We include modes that do not satisfy real analyticity since they can be useful when considering fluctuations of several CDD factors. To understand these $6$ zero modes, it is helpful to consider the following simpler example of S-matrix
\begin{equation}\label{3F}
S_I(\theta)=f(\theta,\alpha_0),\quad S_+(\theta)=f(\theta,\alpha_0),\quad S_-(\theta)=f(\theta,\alpha_0),
\end{equation}
which corresponds to a free theory S-matrix dressed with one CDD factor. Using the change of variables $\theta\rightarrow z$ in eq.(\ref{stheta})
 the CDD factor can be written as
\begin{equation}
f(z,z_0)=\frac{\bar{z}_0}{z_0} \frac{(z-z_0) (z+z_0)}{(z\bar{z}_0-1)(z\bar{z}_0+1)}
\end{equation}
where $z_0$ is the corresponding zero inside the unit disk and $f(z)$ satisfy crossing and saturate unitarity:
\begin{equation}
f(-z)=f(z),\quad |f(e^{i\phi})|=1.
\end{equation}
Now consider a small deformation of the S-matrix $S_a(\theta)\to S_a(\theta)\pm\xi_a(\theta)$ which preserves the crossing symmetry and unitarity saturation for both signs (see fig.\ref{Vdef}). We then have
\begin{equation}\label{xicross}
\xi_a(i\pi-\theta)=C_{ab}\xi_b(\theta)
\end{equation}
and to first order
\begin{equation}
S^*_a(\sigma)\xi_a(\sigma)+S_a(\sigma)\xi^*_a(\sigma)=0.
\label{Sxi}
\end{equation}
A described in Appendix C, finding the analytic functions $\xi_a(z)$ is the so called Riemann-Hilbert problem  that, in this case can be solved by elementary methods. Indeed, since $S^*_a(\sigma)=\frac{1}{S_a(\sigma)}$, we can therefore define the meromorphic functions
\begin{equation}
\eta_a(\theta)=\frac{\xi_a(\theta)}{S_a(\theta)},
\end{equation}
with the property
\begin{equation}
\text{Re}[\eta_a(\sigma)]=0.
\end{equation}
In the case of this simple example \eqref{3F}, the $\eta$'s satisfy the same crossing condition as the $\xi$'s \eqref{xicross}, and have poles where the $f$ has zeros. Therefore, one has
\begin{equation}
\text{Re}[\eta_a(\sigma)]=0,\quad \text{Re}[\eta_a(i\pi-\sigma)]=0.
\end{equation}
In terms of the unit disk in $z$, this corresponds to meromorphic functions with zero real part on the boundary of the unit disk. One can construct such functions explicitly:
\begin{eqnarray}
h_1(z)&=& i\left(\frac{1}{z-z_0}+\frac{1}{\frac{1}{z}-\bar{z}_0}\right),\\
h_2(z)&=&\frac{1}{z-z_0}-\frac{1}{\frac{1}{z}-\bar{z}_0}+\frac{1}{z_0}.
\end{eqnarray}
The last term in the expression of $h_2(z)$ is added to ensure that $h_2(z=0)=0$. We include $h_2(z)$ for completeness but it does not satisfy the real analyticity constraint so it is a deformation outside the space of allowed S-matrices. 
With $h_1$ and $h_2$, we can also write crossing symmetric and antisymmetric combinations:
\begin{eqnarray}
H_{1,2}(z)&=& h_{1,2}(z)+h_{1,2}(-z),\\
\tilde{H}_{1,2}(z)&=& h_{1,2}(z)-h_{1,2}(-z),
\end{eqnarray}
They give rise to the following zero modes for $\{\eta_I,\eta_+,\eta_-\}$:
\begin{eqnarray}
\hat{\eta}_1&=& \{H_1,H_1,H_1\},\\
\hat{\eta}_2&=& \{(N+1)H_1,H_1,-H_1\},\\
\hat{\eta}_3&=& \{(N-1)\tilde{H}_1,-\tilde{H}_1,\tilde{H}_1\},
\label{etah}
\end{eqnarray}
plus another three zero modes obtained replacing $H_1\rightarrow H_2$ giving the $6$ zero modes of \eqref{3F}. It should be noted that only $\hat{\eta}_{1,2,3}$ satisfy real analyticity.

Consider now the case
\begin{equation}\label{ShatF}
S_a(\theta)=f(\theta,\alpha_0) \hat{S}_a(\theta),
\end{equation}
where $\hat{S}_a$ indicate the integrable model S-matrix. Numerically one still finds 6 zero modes. Analytically we were able to identify 4 of them:
\begin{eqnarray}
\xi_{1a}&=& H_1(z(\theta)) f(\theta,\alpha_0) \hat{S}_a(\theta), \\
\xi_{2a}&=& H_2(z(\theta)) f(\theta,\alpha_0) \hat{S}_a(\theta), \\
\xi_{3a}&=& i \tilde{H}_1(z(\theta)) f(\theta,\alpha_0) \frac{\partial \hat{S}_a(\theta)}{\partial \theta}, \\
\xi_{4a}&=& i \tilde{H}_2(z(\theta)) f(\theta,\alpha_0) \frac{\partial \hat{S}_a(\theta)}{\partial \theta}. 
\end{eqnarray}
Again, only two of them are real analytic.

\subsection{Other Vertices}

\begin{figure}
 	\includegraphics[trim=0cm 5cm 0cm 5cm, clip=true, width=\textwidth]{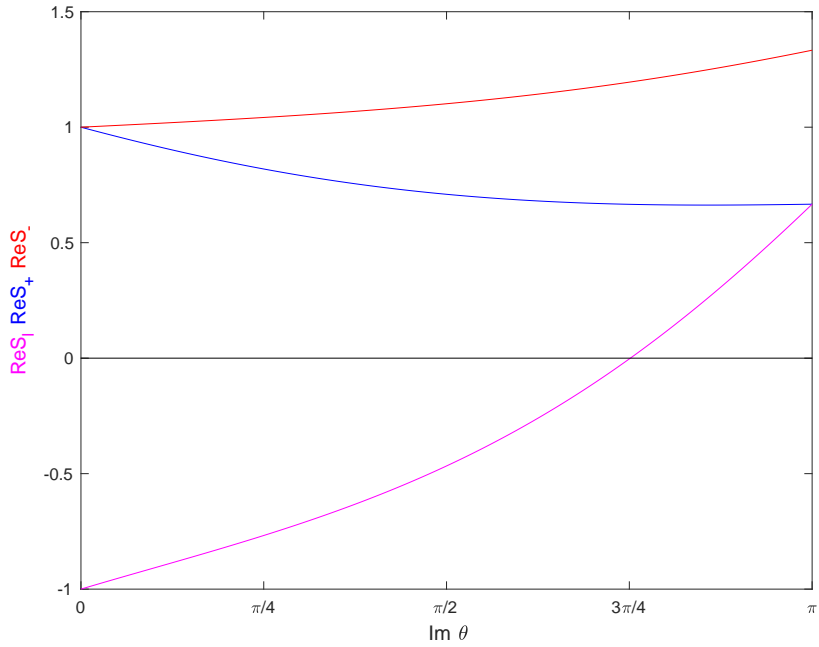}
 	\caption{Other vertex for $N=6$ obtained by maximizing the functional $F = \Re [ S_1(z_0) - \alpha S_2(z_0) ]$ with $\alpha=2.5, z_0=0.3i$. We plot $S_I$ (purple), $S_+$ (blue) and $S_-$ (red) on the imaginary axis of $\theta$, \ie\ $\theta=i\eta$, $0\le\eta\le\pi$. This should be compared with the integrable model, see Figure \ref{imag}. The same vertex can be found for other values of $N$.}
 	\label{otherVertex}
 \end{figure}

 By choosing other linear functionals to maximize, it is possible to find other vertices in the boundary of the space. We give an example in Figure \ref{otherVertex}. This is purely a numerical result, that is, we obtain functions that saturate the constraints and such that the corresponding linearized constraints had no zero modes. It will be interesting if the existence of these vertices can be confirmed by finding their analytic expressions. Since the Yang-Baxter equation leads to the integrable model \cite{Zamolodchikov:1978xm} these vertices should not correspond to integrable theories, at least in the S-matrix factorization sense.  

\section{Conclusions}
 
  It was recently argued that the S-matrix of certain theories can be computed by maximizing a coupling between particles and their bound states under the assumption of a fixed spectrum \cite{Paulos:2016but,Paulos:2016fap,Paulos:2017fhb}.  It was not clear how the program proceeded if the particles formed no bound states. For that reason, in this paper we considered the 2d $O(N)$ bosonic integrable model that is exactly solvable and has no bound states.
  What we found was that there is an infinite number of functionals that give rise to this model by maximization. We emphasize that the maximization procedure reproduces the S-matrix of the integrable model using only the assumptions of crossing, real analyticity and the unitarity bounds. We do not use any extra information about zeros, poles or other properties of the S-matrix. The reason is that this theory has no 
  parameters other than $N$ which appears in the crossing matrix and therefore the S-matrix should be fixed without any extra information. 
  
  Given that there is an infinite number of functionals that one can maximize, instead of concentrating on the functional, we proposed that the defining property of the model is that it is at a vertex of the convex space of allowed S-matrices.
  Then, it is clear that any linear functional whose gradient points in the general direction of the vertex is maximized precisely at the vertex. In fact, maximizing a linear functional in a convex space is a standard numerical problem related to semi-definite programming and solvable by fast algorithms. There has to be some initial trial and error in choosing an appropriate functional but once found, it can be iteratively improved by linearizing the convex space near the vertex.
  We expect this can lead to a general method to define field theories without continuous parameters\footnote{In certain cases, even if the theory has continuous parameters, the same procedure applies if they can be fixed in the S-matrix. One example is the mass of the bound state in the sine-Gordon model as described in Appendix B.}. Such theories should lie at particular points of the boundary of the convex space of allowed S-matrices and should be found by maximizing linear functionals in such space. Of ciurse it will be of great interest if this idea applies to QCD, see the recent work \cite{Paulos:2017fhb,Miro} for interesting ideas in higher dimensions.

\section{Acknowledgments}

 We are very grateful to Luc\'\i a G\'omez C\'ordova and Pedro Vieira for discussions on their related research. We are also grateful to Oleg Lunin and Anatoly Dymarsky for comments and suggestions. This work was supported in part by DOE through grant DE-SC0007884. 
 
{\bf Note:} While this work was being completed we learned of \cite{Cordova:2018uop} that also studies the 2d $O(N)$ non-linear sigma model using the S-matrix bootstrap. After this work was submitted \cite{Paulos:2018fym} appeared with related results.

\appendix
\numberwithin{equation}{section}

\section{Numerical approach}\label{appA}

 The physical region is defined as $0\le \Im \theta \le \pi$ in the $\theta$ plane and as $|z|\le 1 $ in the $z$ plane. The relation between the two is
 \beq
 z = \frac{i-e^\theta}{i+e^\theta}.
 \eeq
  The physical line ($E>2m$) is the real axis in $\theta$ coordinates and $|z|=1, \Im z\ge 0$ in the $z$ plane. By assumption, there are no bound states so the scattering matrices are analytic in the physical region (no poles). 
Given a general, analytic function in the unit disk, the real and imaginary parts are related by
\beq
 \Im\, \phi(e^{i\varphi_0}) = -\frac{1}{2\pi} \int_{0}^{2\pi}\cotan\left(\frac{\varphi-\varphi_0}{2}\right)\,  \Re\ \phi(e^{i\varphi})\ d\varphi\ .
\eeq
This equation becomes eq.(\ref{Kernel}) after the change of variables $\sigma=-\atanh(\cos\varphi)$ which maps the upper half circle in $z$ to the real line of $\theta$.
The first step is to find a numerical analog to this relation, that is, given the values of $\Re \phi(z_j=e^{i\varphi_j})$, $\varphi_j=\frac{j\pi}{M}$ for $j=1\ldots 2M$ we would like to find the approximate values of the imaginary part at those same points, the approximation being better as $M\rightarrow \infty$.
Before going into the derivation let us mention that the result is a straight-forward discretization of the integration kernel in the previous integral:
\beqa
 \Im\, \phi(z_j) &=& \sum_{j'=1}^{2M} K_{jj'} \Re\,\phi(z_{j'}) , \ \ \ \ \ \ z_j =  e^{i\varphi_j}=e^{i\frac{j\pi}{M}}, \label{eqK} \\
  K_{jj'} &=& \frac{(1-(-)^{j-j'})}{2M}  \cotan\left(\frac{\pi}{2M}(j-j')\right), \ \ \ \ \ j\neq j', \\
  K_{jj} &=& 0\ .
\eeqa
 The derivation also gives us the analytic function inside the disk and proceeds as follows. Given the values of 	$\Re \phi(z_j)$ we approximate the real part at the boundary of the disk as
\beq
 \Re \phi(z_j) = \sum_{j=1}^{2M} d^{(M)}(\varphi-\varphi_j)\, \Re \phi(z_j)\ ,
\eeq 
 where
\beq
 d^{(M)}(\varphi) = \frac{1}{2M}  \frac{\sin M\varphi}{\tan \frac{\varphi}{2}}\ .
\eeq	
 Now the task is to analytically continue $d^{(M)}(\varphi)$ inside the unit disk obtaining the function
\beq
 \hat{d}^{(M)}(z) = \frac{1}{2M} \frac{1+z}{1-z}(1-z^M)
\eeq
which is analytical and satisfies
\beq
\Re\ \hat{d}^{(M)}(z=e^{i\varphi}) = d^{(M)}(\varphi) = \frac{1}{2M} \frac{\sin M\varphi}{\tan \frac{\varphi}{2}}\ .
\eeq
 Thus, our best guess for the analytic function $\phi(z)$ is 
\beqa
 \phi(z) &=& \sum_{j=1}^{2M} \hat{d}^{(M)}(\frac{z}{z_j})\, \Re \phi(z_j) \\
           &=& \frac{1}{2M} \sum_{j=1}^{2M} \frac{z_j+z}{z_j-z}\left[1-\left(\frac{z}{z_j}\right)^M\right]\, \Re \phi(z_j)\ ,
\eeqa
from where we can evaluate the imaginary part at the boundary of the disk giving \eqn{eqK} and also evaluate the function in the interior as needed for maximization.

\section{The sine-Gordon model}
 In this appendix we briefly review the results of \cite{Paulos:2016fap,Paulos:2016but,Paulos:2017fhb} and show that the idea of the theory at a vertex also applies in that case. The S-matrix in which we are interested describes the scattering of two particles of mass $m$ associated with the scalar field of sine-Gordon. In this channel there is a bound state of mass $m_1 = 2m\sin\frac{\gamma}{16}$. Here $\gamma$ is a real parameter $4\pi\le \gamma\le 8\pi$. In the $s$ variable the S-matrix has poles corresponding to the bound state in the $s$ and $t$-channel and reads
 \beq
  S(s) = -\frac{\tilde{g}}{s-m_1^2} - \frac{\tilde{g}}{4m^2-s-m_1^2} + \tilde{S}(s)\ ,
 \eeq
where $\tilde{g}$ can be understood as a coupling constant between the particle and the bound state and $\tilde{S}(s)$ is an analytic function in the physical $s$-plane with cuts on the real axis for $-\infty<s\le0$ and $4m^2\le s<\infty$.
 The S-matrix is more easily written in the $z$ plane as
 \beq
  S(z) = \frac{ig}{z+ia}-\frac{ig}{z-ia} + \hat{S}(z), 
 \eeq
 with
 \beq
 a = -\frac{\cos(\frac{\gamma}{8})}{1+\sin(\frac{\gamma}{8})} >0 , \ \ \ g=\frac{\tilde{g}}{4m^2}\frac{(1+a^2)^2}{1-a^2}\ ,
 \eeq
and $\hat{S}(z)$ a holomorphic function in the unit disk such that $\hat{S}(z)=\hat{S}(-z)$ by crossing and $\Im \hat{S}(i\xi) =0$ for $-1\le\xi\le1$ by real analyticity. Consider now the CDD factor
\beq
 f_0(z) = \frac{z^2+a^2}{1+a^2z^2}
\eeq
 such that $|f_0(e^{i\phi})|=1$, $f_0(z)=f_0(-z)$ and $f_0$ is analytic in the unit disk with zeros at $z_\pm=\pm ia$. Define now
\beq
S_1(z) = f_0(z) S(z)  
\eeq
It is clear that $S_1(z)$ is analytic in the unit disk, and satisfies $|S_1(e^{i\phi})|\le 1$, crossing and real analyticity. Moreover
\beq
 S_1(ia) = 2g\frac{a}{1-a^4}
\eeq
which is real as required by real analyticity. In \cite{Paulos:2016fap,Paulos:2016but,Paulos:2017fhb} it was proposed to maximize $\tilde{g}$ or, as we see here, equivalently $\Re(S_1(ia))$. Since the maximum of the modulus of $S_1(z)$ should be at the boundary of the disk, and is bounded by $|S_1(e^{i\phi})|\le 1$, the maximum is  
$\mbox{max}[\Re(S_1(ia))]=1$ attained by the constant function $S_1(z)=1$. This gives
\beq
 S(z) = \frac{1}{f_0(z)} = \frac{1+a^2z^2}{z^2+a^2} = \frac{\sinh\theta + i \sin\frac{\gamma}{8}}{\sinh\theta - i \sin\frac{\gamma}{8}}\ ,
\eeq
which agrees with the $S^{(1,1)}(\theta)$ scattering matrix in the sine-Gordon model \cite{Zamolodchikov:1978xm} as already explained in \cite{Paulos:2016but}. Here we want to point out that we could equally well maximize
\beq
 F_{\mbox{max}} = \Re(S_1(ib))
\eeq
for any $-1<b<1$ and the maximum will be the same $S_1(z)=1$, \ie\ there is an infinite number of functionals that we can maximize. Since
\beq
 S(ib) = \frac{1-a^2b^2}{a^2-b^2} S_1(ib)\ ,
\eeq
we can equivalently write the functional $F_{\mbox{max}}$ in terms of the original S-matrix $S(z)$:
\beq
 F_{\mbox{max}} = \Re(S(ib)), \ \ \mbox{for}\ b^2<a^2 ,\ \ \ \mbox{or}\ \ F_{\mbox{max}} = -\Re(S(ib)), \ \ \mbox{for}\ b^2>a^2 \ .
\eeq
 For the purpose of this paper it is useful to check if this S-matrix is at a vertex of the allowed space. Consider a fluctuation
\beq
 S_1(z) = 1 \pm \xi(z)\ ,
\eeq 
with $\xi(z)$ analytic in the unit disk and satisfying crossing and real analyticity: 
\beqa 
\xi(z)&=&\xi(-z),\\
\Im(\xi(i\eta)) &=& 0,\ \  -1\le\eta\le 1 .
\eeqa 
Since we require $|S_1(e^{i\phi})|\le1$, at first order for both signs $\pm\xi$ (see fig.\ref{Vdef}) we should have
\beq
 \Re(\xi(e^{i\phi})) =0\ .
\eeq
 This implies that $\xi(z)=0$ and therefore there are no zero modes. The integrable model is once again at a vertex of the space of allowed S-matrices.

\section{Relation to the Riemann-Hilbert problem}

 In section \ref{CDD} we discussed that the problem of finding fluctuations to the maximum is equivalent to the RH problem. As discussed in eq.(\ref{Sxi}), what we want to find is a fluctuation $\xi_a(z)$ analytic and crossing symmetric such that 
\beq
 S_a(t) \xi^*_a(t) + S^*_a(t) \xi_a(t) =0 , \ \ \ \ t=e^{i\phi}, \ \ \ 0<\phi<\pi 
 \label{Sxi2}
\eeq
Due to crossing symmetry we also have 
\beq
  S_a(-t) \sum_b C_{ab} \xi^*_b(t) + S^*_a(-t) \sum_{b} C_{ab} \xi_b(t) =0 , \ \ \ \ t=e^{i\phi}, \ \ \ \pi<\phi<2 \pi 
\eeq
To obtain the standard formulation of the RH problem \cite{singeq}, define functions analytic inside the disk $\xi_a^{(+)}(z)=\xi_a(z)$ and others outside the disk as
\beq
 \xi^{(-)}_{a}(z)=\left[\xi_a\left(\frac{1}{\bar{z}}\right)\right]^*
\eeq
such that 
\beqa
 \xi_a^{(+)}(t)&=&\lim_{z\rightarrow t,|z|<1} \xi_a(z) = \xi_a(t) \\
\xi_a^{(-)}(t)&=&\lim_{z\rightarrow t,|z|>1} \left[\xi_a\left(\frac{1}{\bar{z}}\right)\right]^*=  (\xi_a(t))^* 
\eeqa
If we also define the matrix
\beq
 \tilde{C}_{ab}(t) = \left\{\begin{array}{ccl}
 	\hat{S}^*_{a}(t)\, \delta_{ab}&\ \ \ \   & 0<\phi<\pi \\
 	\hat{S}_a^*(-t)\, C_{ab} && \pi<\phi<2\pi
 \end{array}\right.
\eeq
then the problem requires finding two analytic functions, $\xi^{(\pm)}_a$, one outside the disk $\xi^{(-)}_a(z)$ and the other one inside $\xi^{(+)}_a(z)$, such that their boundary values are related by 
\beq
 \xi_a^{(+)}(t) = -\tilde{C}_{ab}^{-1}(t)\, \tilde{C}^*_{bc}(t)\, \xi_c^{(-)}(t).
\eeq
 Given such functions, the solution that satisfy the condition \eqref{Sxi2} and crossing can be easily constructed as
\beq
 \xi_a(z) = \frac{1}{4}\left(\xi^{(+)}_a(z)+\left[\xi^{(-)}_a\left(\frac{1}{\bar{z}}\right)\right]^*+C_{ab}\xi^{(+)}_b(-z)+C_{ab}\left[\xi^{(-)}_b\left(-\frac{1}{\bar{z}}\right)\right]^*\right)
\eeq
If the RH problem has no solutions then the theory is at a vertex of the space of S-matrices, if not, there is a continuum of theories satisfying the constraints in the vicinity of the maximum. Unfortunately, the vector RH problem has no analytical solutions, numerical solutions are equivalent to checking the zero modes of the matrix $V$ as previously discussed around eq.(\ref{VTV}). In the case of the simple S-matrix (\ref{3F}) the RH problem is diagonal and can be solved by standard methods \cite{singeq}. The result is 
\beq
 \xi^{(+)}_a = \frac{1}{(z^2-\frac{1}{\bar{z}_0^2})^2} \sum_{\ell=0}^{4} \alpha_{a\ell} z^\ell 
\eeq
After imposing crossing we obtain
\beq
 \xi_a = \frac{(1+z^4)(\alpha_1\psi_{1a}+\alpha_2\psi_{2a})+z^2(\alpha_4\psi_{1a}+\alpha_5\psi_{2a})+(\alpha_3 z+\alpha_3^* z^3)\psi_{3a}}{(z^2-\frac{1}{\bar{z}_0^2})^2}
\eeq
which agrees with the simpler calculation in eq.(\ref{etah}). Here the constant vectors $\psi_{1,2,3}$ are the eigenvectors of the crossing matrix $C$:
\beq
 \psi_1=\left(\begin{array}{c}1\\ 1\\ 1\end{array}\right), \ \ \ 
 \psi_2=\left(\begin{array}{c}\frac{N}{2}\\ 0\\ -1\end{array}\right), \ \ \ 
 \psi_3=\left(\begin{array}{c}N-1\\ -1\\ 1\end{array}\right), 
\eeq
with eigenvalues $1,1,-1$ respectively.

\clearpage
\newpage

{\bf Addendum:}  Matlab program for the $O(6)$ model using the cvx package. It plots the real part of $S_{I,+,-}$ on the real axis of the $\theta$ plane.
\begin{lstlisting}
% Matlab program, requires cvx package
% O(6) model with M=200 interpolating points. Iterative improvement.
N = 6;  M = 200;	
% constructing the K matrix
vc = 1/M*cot(pi/2/M*[1:2*M-1]); 
vc(2:2:end) = 0; 
Km = toeplitz([0,vc],zeros(2*M,1)); 
Km = Km - Km.'; 
% Crossing symmetry matrix
C = [1/N N/2+1/2-1/N (1-N)/2; 1/N 1/2-1/N 1/2; -1/N 1/2+1/N 1/2];   
K = kron(eye(3),Km(1:M,1:M))+kron(C,Km(1:M,(M+1):end)); 
for count = 1:10         % 10 iterations
 cvx_begin quiet 
  variable ReS(3*M)
  ImS = K*ReS;
  ReS.*ReS + ImS.*ImS <= 1;
  % evaluating the functions at z0 
  z0 = 0.3*i;
  zj = exp(1.i*pi/M*[1:M]);
  w1 = 1/2/M*(1-(z0./zj).^M).*(zj+z0)./(zj-z0);
  w2 = 1/2/M*(1-(z0./zj).^M).*(zj-z0)./(zj+z0); 
  W0 =  kron(eye(3),real(w1)) + kron(C,real(w2)); 
  v1 = 1/2*(W0(2,:)+W0(3,:));
  v2 = (1/2/N*W0(1,:)+(1/4-1/2/N)*W0(2,:)-1/4*W0(3,:));
  if count==1   t = (v1-7.5*v2)*ReS;  
  else              t = w0*ReS;                  
  end
  maximize(t);
 cvx_end
 V = ReS.*eye(3*M)+ImS.*K;  % infinitesimal variations around maximum
 w0 = ones(1,3*M)*V;     % new maximization functional
end
sigma = -atanh(cos(pi/M*[1:M-1]));
plot(sigma,ReS(1:M-1),'o',sigma,ReS(M+1:2*M-1),'+', ...
      sigma,ReS(2*M+1:3*M-1),'d')
\end{lstlisting}

\end{document}